\documentclass[sigconf]{acmart}

\usepackage{csquotes}

\acmConference[P-RECS 2018]{First International Workshop on Practical Reproducible Evaluation of Computer Systems.}{June 2018}{Tempe, Arizona USA}
\acmYear{2018}
\copyrightyear{2018}

\begin{document}
\title{Applying Distributed Ledgers to Manage Workflow Provenance}

\author{Jay Jay Billings}
\authornote{Corresponding author. Email or Twitter: @jayjaybillings}
\orcid{0000-0001-8811-2688}
\affiliation{%
  \institution{Oak Ridge National Laboratory}
  \streetaddress{PO Box 2008 MS6173}
  \city{Oak Ridge}
  \state{TN}
  \postcode{37830}
  \country{USA}}
\email{billingsjj@ornl.gov}

\begin{abstract}
Sharing provenance across workflow management systems automatically is not currently possible, but the value of such a capability is high since it could greatly reduce the amount of duplicated workflows, accelerate the discovery of new knowledge, and verify the integrity of past and present analyses. Although numerous technological challenges exist to efficiently share provenance information across workflow management systems, permissioned distributed ledgers could surmount many of them. The primary benefit of permissioned distributed ledgers over other technologies is that their distribution is over a peer-to-peer network that encodes transactions across the network into an immutable hash list and achieves consensus on the validity of the new data through a common consensus mechanism. This work discusses provenance and distributed ledgers on their own and then presents an argument that distributed ledgers naturally satisfy many of the requirements of workflow provenance, that provenance information can exist in the ledger in multiple ways, and that a number of novel research areas exist based on this strategy.
\end{abstract}


\maketitle

\underline{Notice of Copyright:} This manuscript has been authored by UT-
Battelle, LLC, under contract DEAC05-00OR22725 with the US Department of
Energy (DOE). The US government retains and the publisher, by accepting
the article for publication, acknowledges that the US government
retains a nonexclusive, paid-up, irrevocable, worldwide license to publish or
reproduce the published form of this manuscript, or allow others to do so, for
US government purposes. DOE will provide
public access to these results of federally sponsored research in accordance
with the DOE Public Access Plan (http://energy.gov/downloads/doe-public-access-plan).

\section{Introduction}
Consider the following statement about the provenance of a piece of data $A$:
\begin{displayquote}
    $A$ was generated using linear regression on $B$ and stored in file $C$ yesterday.
\end{displayquote}
Provenance capture as it pertains to information, in this case $A$, is the process of collecting metadata about the creation, manipulation, and use of that information. This is important when the history of $A$ needs to be trusted and verified or in situations where $A$ needs to be reused or reproduced reliably. The provenance in this case describes how $A$ was generated (linear regression), what data it was generated from ($B$), where it was stored ($C$), the nature of that storage (a file), and when these things occurred (yesterday). 

The previous example could also be taken as a description of a very basic scientific workflow describing how to manipulate $B$ to generate $A$ and stage the results in file $C$. The overlap is this close only in the most basic examples. One key difference between these perspectives is that the workflow description is executable in some workflow management system, but the provenance is a persistent record about that workflow. Over time the provenance trail for $A$ might also encompass workflows across many executions or multiple workflow management systems and have additional preservation requirements. This conceptual proximity leads many workflow management systems to include complete workflow descriptions in provenance records, and some workflow management systems can recreate and execute workflows from their provenance records alone.

However, one significant problem for provenance as it relates to scientific workflows it that there are no standard ways to share it. Although some workflow management systems provide integrated provenance repositories, the practice is not universal and the provenance repositories might not be designed with other systems in mind. What if a highly cited publication relied on results generated from a scientific workflow that could be reproduced quickly if its full history was available? Furthermore, what if it were possible to compute new and equally significant results in record time by using the history to generate a new workflow description with minor modifications to the original? Both of these may be relatively simple to accomplish in a single, well-designed workflow management system that captures provenance, but there is no readily available general-purpose solution.

This work argues that a \textit{permissioned distributed ledger} is an efficient solution to share workflow provenance between networked peers. The distributed ledger would act as the broker between peers to execute workflows as transactions on the network, and each transaction would be recorded in the ledger as usual. The act of recording the transactions would, in effect, be capturing the root node of the provenance graph, which would be augmented by additional information stored in the transaction record to point to system-specific provenance information. The benefits to this are that transactions in distributed ledgers are (1) shared across the network to establish consensus on their validity and (2) made immutable by encoding the state in a distributed hash list, which insures integrity. Trust among peers is established by cryptographic keys in permissioned ledgers (as opposed to proof of work schemes), so only minor resources are required to join the network. 

The remainder of this article investigates this possibility. Section \ref{background} provides a general overview of provenance as well as the basics behind distributed ledgers. Section \ref{case} presents the case for using distributed ledgers for this purpose in more detail.

\section{Background}
\label{background}

\subsection{Workflow Provenance}
The issues around provenance for workflows, including scientific workflows, have been thoroughly investigated elsewhere in the literature. Data provenance for workflows has been investigated particularly thoroughly in the data science community, (c.f. \cite{davidson_provenance_2007}). Workflow management systems for modeling and simulation capture provenance as well, but some do so by explicitly generating graph-based provenance \cite{pizzi_aiida:_2016}, while others follow a ``log everything'' style \cite{billings_eclipse_2017}. 

Multiple provenance standards exist, most of which are applicable to problems in broader scientific domains as well as workflow science. The PROV Model, for example, is a standard for capturing provenance developed as a successor to the Open Provenance Model \cite{noauthor_prov-overview_nodate} \cite{moreau_open_2011}. In PROV, provenance is modeled with a data model, PROV-DM, and can be stored in XML, RDF, Dublin Core, and a human-readable form for examples. Constraints can also be applied using the PROV-CONSTRAINTS module. Provenance is stored in a tree in PROV, and the data model includes entities, activities, usage, generation, time, and other elements. 

In tree-based provenance models such as PROV, provenance is captured in a tree beginning with an element, usually an entity, at the root node. Entities are normally the root node because they are often the end product or are used as the source to generate other data. Depending on the model, the element at the root node will be either the initial element or the final entity. For example, PROV roots the provenance tree in the final created entity, and edges in the tree indicate usage or generation going backwards in time from the end to the beginning. This is a natural representation when considering the question ``Where did this element come from?''; whereas a time-forward perspective is natural when considering the question ``What did this element become or produce?'' Activities in tree-based provenance models are models as nodes, like entities.

Event-based or ``log everything'' provenance models provide detailed provenance information based on logs created from events in the system. Workflow management systems can easily generate this type of provenance trail as the workflow is executed (which is why it was chosen in \cite{billings_eclipse_2017}). As the workflow is executed, the input and output are captured at each step, as well as a description of or the entire instruction set, and logged. The entire workflow description might be saved as well. Information on activities, entities, time, and other elements can be captured just like a tree-based provenance model. The provenance information is most commonly available in the form of logs but might be generated as separate provenance reports in some systems. Although it is conceivable that a provenance trail in an event-based system could be ordered from with time running backwards, like all event-based systems it is much more common to find a time-forward representation. Some systems, including those in \cite{billings_eclipse_2017} and \cite{altintas_provenance_2006}, can use this type of provenance record to enable ``fast replays'' of workflows.

\subsection{Distributed Ledgers}
Distributed ledgers are linked collections of records about transactions that are distributed across a peer-to-peer network without any central authority. The Blockchain data structure that forms the basis of the BitCoin cryptocurrency is the most well-known implementation of a distributed ledger \cite{nakamoto_bitcoin:_nodate}. Records, or groups of records called Blocks, are linked in order through a hash list where each item is linked to the one before and after it through a unique hash of the record(s), forming a ``chain.'' Without a central authority to certify the validity of records, the means of determining consensus on whether or not records are valid requires the use of a consensus algorithm executed by nodes in the network. The exact algorithm used depends on whether or not the ledger is open or permissioned. Open networks, such as cryptocurrencies, tend to determine consensus through proof-of-work algorithms \cite{nakamoto_bitcoin:_nodate} or through proof of stake \cite{noauthor_proof--stake_2018}.

The basic operation of a distributed ledger is as follows, assuming an underlying Blockchain implementation:
\begin{itemize}
    \item A transaction is executed on the network between entities $A$ and $B$ to create an asset $C$.
    \item The transaction is time-stamped and added to a collection of unverified transactions, which are linked to previously verified transactions through a hash.
    \item A check (such as proof of work) is executed on the collected transactions to ascertain their validity. 
    \item The majority of nodes in the remainder of the network comes to a consensus on the result of the check as presented by the original node or nodes that checked the collection.
    \item The collection is accepted, and its hash is used for the next collection; the process is then repeated.
\end{itemize}

The most appealing features of distributed ledgers are that a central authority is not required and that the combination of consensus algorithms and a hash list (or hash tree) to verify and store the transactions creates a very reliable system. Because of this, distributed ledgers have found application in cryptocurrencies, traditional financial markets, and many other areas \cite{noauthor_groups:requirements:use-case-inventory_nodate} \cite{noauthor_discussion_2018}. The applicability of Blockchain to business process modeling has also been investigated \cite{Mendling:2018:BBP:3146385.3183367}.

\subsubsection{Permissioned versus Open Networks}
Open networks must use proof-based algorithms to establish trust because members of the network are inherently untrustworthy. Thus, by providing proof that is acceptable to a majority of the remaining network, nodes can be added to the list. Since the proof might be computationally expensive, cryptocurrencies provide the incentive of receiving coins in the currency appropriate to the work.

If the primary motivation of such a costly proof scheme is to create trust between untrustworthy parties, one obvious alternative is to work only with trustworthy nodes. Permissioned networks are formed by nodes identified by strong cryptographic keys and allowed to join by permission of other members in the network. Reaching consensus in this situation is as simple as making sure that the source of the transaction and its purpose are valid between some parties on the network. No incentive is required to check transactions and vouch for them in this case beyond membership in the network, which improves performance and alleviates any concern about the work, stake, or cost required to participate.

\subsubsection{Smart Contracts}
A smart contract, or simply a contract, in this context is a small piece of code that is executed in response to a transaction. Business logic executed on the network is done so through contracts. Contracts can have many uses, although some ledgers may limit the type of code that can be executed for either architectural reasons or security.

\subsubsection{Relationship to other technologies}
There are number of technologies closely related to distributed ledgers. Blockchain data structures are implemented using Merkle trees \cite{merkle_digital_1987} as are other distributed databases and version control systems, such as Git. There is also substantial on-going work to improve on the performance of distributed ledgers and Blockchains in particular. For example, the PHANTOM protocol, based on BlockDAG, alleviates many of the performance issues that are side effects of Nakamoto's original consensus scheme and allows for asynchronous, fast block creation. Other efforts are also investigating the use of Blockchain technologies for provenance capture \cite{richard_brooks_and_anthony_skjellum_using_2017}\cite{worley_scrybe:_nodate}.

\section{Distributed Ledgers for Workflow Provenance}
\label{case}
The promise of a permissioned ledger for managing workflow provenance is that a large, secure network of peers can quickly and automatically share provenance information without additional work and in a way that preserves the integrity of the provenance record. Although there might be other means to do accomplish this task, permissioned distributed ledgers have additional properties that map well onto those required of a good provenance record.
\begin{itemize}
    \item Transaction records, which would describe workflow executions in this case, are immutable. Placement in the ledger requires linking to the previous and next records in the Merkle tree/hash list/Blockchain backing the ledger. Thus the record cannot be changed without changing the entire set of records.
    \item Transaction records are secure. In addition to inheriting immutability, which is itself a good component of security, records are deemed valid by trusted network peers. This offers both security through consensus and security through the ``reputation'' of the peers.
    \item Transaction records can hold significant amounts of state. Records include information about themselves, such as when they were created, what the transaction was meant to do, etc. However, contracts can be used to inject additional state into the records, which could include root node provenance information, or to provide extra provenance parameters to the workflow management system on the supply side of the transaction.
    \item Transaction records are \textit{uniquely} identifiable in the ledger and can be found through queries.
    \item Transaction records are \textit{uniformly} identifiable in the ledger, meaning that the same method to identify one transaction can be used to uniquely identify others. Thus, at a very high level, such a scheme would imply that the root of any provenance record could be found without requiring knowledge of the exact provenance standard used by the workflow management systems at the nodes.
    \item The ledger can be walked easily in either time-backward or time-forward order, matching either tree-based or event-based provenance models.
\end{itemize}

These properties also suggest that ledgers would be good for managing provenance regardless of whether or not the ledger was distributed. (Trusted peers could be queried secretly and their consensus computed locally, for example.)

\subsection{Capture Model}
In a distributed ledger, workflow management systems would be peers on the network, along with clients that represent humans and possibly other service nodes, such as nodes for staging data. The latter point makes sense because moving data can itself be considered a part of setup or postprocessing workflows. Transactions---workflow executions---would occur when one peer in the network asks another to execute a workflow, which would start by executing a contract to initiate the actual workflow execution. Once the task was complete, the transaction would be tested for validity and consensus gained on the network, at which point details about the workflow execution would be logged. Workflow management systems could either export the entire provenance record into the contract code that executed the workflow, which could write it as state in the network, or add an entry to the state that describes the provenance standard used by the system and a second entry pointing to the file/resource that contains the rest of the provenance record. The first case has the benefit that the provenance record will never be lost as long as the ledger exists, but the second case might be more scalable because it requires less compute cycles and stores less data. In practice, these options could be used together without an obvious downside.

\subsection{Access}
There is little reason to consider the use of a distributed ledger for workflow provenance that is not permissioned since presumably all of the peers using the network would be at scientific institutions of one form or another. Thus the numerous issues associated with proof schemes, be they proof of work, stake, or time, could be avoided by registration alone. The full transaction log could be shared publicly as well so that others could benefit from the work, even if they do not contribute to it. However, the question of proprietary data in the ledger requires a slightly different solution.

Proprietary data or data that cannot otherwise be revealed to the public can be stored in a distributed ledger and is not necessarily a reason to choose either an open or permissioned ledger for scientific workflows. Hyperledger, for example, solves the issue of proprietary network in the ledger by introducing \textit{channels} \cite{noauthor_hyperledger_nodate}. Private transactions are handled in private channels, and although the existence of the transaction is recorded in the ledger, no state is saved for that transaction, which hides all the secret information. This is a nice alternative that makes it possible to share some information while saving other information instead of completely hidding the ledger behind permissioning or adopting another technology.

\section{Discussion and Conclusion}
The goal of this work was to demonstrate compatibility between provenance models for scientific workflows and distributed ledgers. There might be significant benefits to using distributed ledgers for capturing provenance because of how easily it enables sharing and search while also mixing in some standardization.

One key question that remains is how well provenance captured in a distributed ledger can be used to quickly create new workflows based on the original. Duplicating a transaction is a relatively straightforward thing and, in principle, executing it might be straightforward as well. However, it might be necessary to investigate how the new transaction can point to the original transaction from which it was copied. Would it be as simple as adding a secondary hash in the saved state that pointed to the parent? Or would it require a ``blocktree'' that allowed branching instead of solely a blockchain/hash list? In any case, the identity of the parent would seem to be important provenance itself!

Another important question is how deep the provenance tree could go. Investigating the degree to which contracts could convert provenance information to standard ledger state is an interesting topic for further research since it would address compatibility issues between workflow and provenance models.

Addressing these and other questions as part of an extensive pilot project would be valuable. There are many open source distributed ledger frameworks, including the previously mentioned Hyperledger, and a number of commercial solutions that could keep the investigation focused on the science rather than extensive software engineering. Success in this endeavor could create an entirely new way of producing reproducible and reusable scientific workflows.

\section*{Acknowledgments}\label{acknowledgments}
\addcontentsline{toc}{section}{Acknowledgments}

The author is grateful for the information and citations on Scrybe from Anthony Skjellum, Director of the SimCenter at the University of Tennessee, Chattanooga. The author is also grateful to the internal reviewers and helpful colleagues at Oak Ridge National Laboratory (ORNL).

This work has been supported by the US Department of Energy, the ORNL Director's Research and Development Fund in the Integrated Computational Environment for the Modeling and Analysis of Neutrons (ICEMAN) project, and by the ORNL Undergraduate Research Participation Program, which is sponsored by ORNL and administered jointly by ORNL and the Oak Ridge Institute for Science and
Education (ORISE). ORNL is managed by UT-Battelle, LLC, for the US Department of Energy under contract no. DE-AC05-00OR22725. ORISE is managed by Oak Ridge Associated Universities for the US Department of Energy under contract no. DE-AC05-00OR22750.

\bibliographystyle{ACM-Reference-Format}
\bibliography{phdThesis}

\end{document}